\documentclass[aps,pra,eqsecnum,twocolumn]{revtex4}
\usepackage{graphicx}
\usepackage{bm}
\usepackage{amsmath,amssymb,amsthm,dsfont,bm}
\usepackage{color}
\usepackage{wasysym}
\usepackage{physics}
\usepackage{ulem}
\usepackage{appendix}
\usepackage[dvipsnames]{xcolor}

\usepackage[english]{babel} %Español 
\usepackage[utf8]{inputenc} %Para poder poner tildes
\usepackage{hyperref}
\usepackage{cancel}

\begin{document}
\preprint{}
\title{Conditional probabilities in quantum-optical settings}
\author{Alfredo Luis}
\affiliation{Departamento de \'Optica, Facultad de Ciencias F\'{\i}sicas, Universidad Complutense, 28040 Madrid, Spain}

\date{\today}

\begin{abstract}
We examine conditional probabilities derived from the joint noisy measurement of two complementary observables. We show that conditioning on one variable can lead to reduction of uncertainty in the other one, even completely eliminating uncertainty. We show that the conditional distribution cannot, in general, be represented in Born form using the marginal positive operator valued measure and any physical system state, nor can it be derived from typical rules of state reduction. We examine these issues in two basic quantum-optical schemes. These are double homodyne detection and a qubit measurement. 
\end{abstract}

\maketitle

\section{Introduction}

The quantum theory is of pure statistical nature. This is well reflected by the fact that nonclassical states are defined to be the ones that lack a joint probability distribution for incompatible observables \cite{MW95,SZ97}. This is actually at the root of the violation of Bell-like tests that confirm the quantum nature of reality \cite{JB64,AF82,SGH21,SH22,AR15,MA84,AA24,DGG05,AP00,AK00,HP04,AM08,TN11,AK14,JCh17}.

In this work we will examine some statistical features related to complementarity. To this end we will consider the noisy joint measurements of incompatible observables leading to legitimate joint distributions conveying complete information about the exact statistics of the original observables \cite{MM90,WMM98,WMM93,WMM02,PB87,PB86,YLLO10,JM68,TH16,AL16,LM17,MAL20,MAL22,AL25a,AL25b,AL25c}. We will focus on conditional probabilities that can be derived from the joint measurement framework. We think that conditional probabilities have an interesting role to play in the proper understanding of the nature of the quantum theory \cite{AK14,IG13,BL09,MO85,PL22}.

Conditional probabilities provide us with the probability distribution for one variable provided that some other one takes a definite value. In quantum mechanics one knows about the value of a variable only once it is measured. So we expect that conditional probabilities may reveal details of the peculiarities of quantum measurement, including complementarity and state reduction. We also would like to examine whether conditional probabilities follow the same Gleason-Born theorem regarding how probabilities are constructed in the quantum realm. We have already obtained some interesting results along these lines as presented in Ref. \cite{PL22}.

As cases of study for these objectives we focus on two typical quantum-optical settings. One concerns qubits, that may be represented in practical terms by the polarization of a single photon. The other one deals with the simultaneous noisy observation of two conjugated quadratures by the double homodyne technique, which is a fundamental quantum-optical measurement scheme  \cite{AK65,CLG87,LP93a,LP93b,KL08}. 

A key feature of joint measurements is that they may include correlations between observables that are essentially forbidden, or even meaningless, in the standard formulation of the quantum theory. These correlations do not enter in the information content of the two observables measured, but they have consequences such as the ones analyzed in this work. 

\section{Squeezed Q function} 

We consider the noisy joint measurement of two conjugate one-mode field quadratures, $X$ and $Y$, represented by the positive operator valued measure (POVM)  
\begin{equation}
\label{POVM}
    \Delta (x,y) = \frac{1}{\pi} D(x,y) \rho_0 D^\dagger (x,y)  
\end{equation}
where $\rho_0$ is a rotated squeezed vacuum, and $D(x,y)$ is the displacement operator 
\begin{equation}
D(x,y) = \exp\!\left[\,2i\left(yX-xY\right)\right] ,
\end{equation}
working under the convention that $[X,Y] = i /2$. The resulting probability distribution is 
\begin{equation}
    \tilde{p} (x,y) = \mathrm{tr} \left [ \rho \Delta (x,y) \right ] .
\end{equation}

This is to say that the statistics of the measurement is actually a squeezed version of the Husimi Q function \cite{KH40,CG69}. The properties of the squeezed vacuum may be well expressed by its Gaussian Wigner function $W_0 (x, y) $ \cite{EW32,WPS01}
\begin{equation}
W_0 (x, y) = \frac{1}{2\pi\sqrt{\det\Sigma}} \exp \left[
-\frac{1}{2} (x,y)\Sigma^{-1} \binom{x}{y} \right],
\end{equation}
where the covariance matrix is
\begin{equation}
\Sigma= \frac{1}{4} R(\theta) \begin{pmatrix} e^{-2r}&0\\ 0&e^{2r} \end{pmatrix} R^T(\theta) = \begin{pmatrix} \Sigma_{xx}&\Sigma_{xy}\\ \Sigma_{yx} & \Sigma_{yy} \end{pmatrix},
\end{equation}
with $R(\theta)$ a rotation matrix in the quadrature plane. The matrix elements of $\Sigma$ are  
\begin{eqnarray}
\label{S}
& \Sigma_{xx}= \frac{1}{4} (e^{-2r}\cos^2\theta+e^{2r}\sin^2\theta), & \nonumber \\
 & & \nonumber \\
& \Sigma_{xy}= \Sigma_{yx}=\frac{1}{4} (e^{2r}-e^{-2r})\sin \theta \cos \theta ,  & \\
 & & \nonumber \\
& \Sigma_{yy}=\frac{1}{4}(e^{-2r}\sin^2\theta+e^{2r}\cos^2\theta) . & \nonumber
\end{eqnarray}
Rotated squeezed states belong to the generalized class of minimum uncertainty states sometimes referred to as intelligent states, \cite{ACS76,DAT94,DAT00}
\begin{equation}
\label{mS}
    \Sigma_{xx}\Sigma_{yy}- \Sigma_{xy}^2 = \frac{1}{16} .
\end{equation}
The variances $\Sigma_{xx}, \Sigma_{yy}$ depending on the squeezing factor $r$ and the rotation angle $\theta$ determine how the unavoidable additional noise is distributed between the $X,Y$ observables. Meanwhile, the term $\Sigma_{xy}$ introduces correlations between them. As simple particular examples we have that if $r=0$ both quadratures would be observed with the same amount of additional noise $\Sigma_{xx}= \Sigma_{yy} = 1/4$ and without induced correlations $\Sigma_{xy}=0$. On the other hand, if $r \neq 0$ and $\theta = 0$ the noise is unequally distributed, $\Sigma_{xx}= e^{-2r}/4$, $\Sigma_{yy} = e^{2r}/4$, again without correlations $\Sigma_{xy}=0$. Later on we will comment on the experimental realization of this setting via double homodyne detection.

\section{Joint statistics} 

The statistics $\tilde{p}(x,y)$ of this measurement can be readily computed using the Wigner function formalism in the form
\begin{equation}
\label{tp}
\tilde{p} (x,y) = \int dx^\prime dy^\prime W_S (x^\prime,y^\prime ) W_0 (x^\prime - x,y^\prime - y ) , 
\end{equation}
where $W_S(x,y)$ is the Wigner function of the system state $\rho$. We can see that the Wigner function of the squeezed vacuum $W_0 (x,y)$ acts as a kernel connecting $W_S(x,y)$ with $\tilde{p} (x,y)$.

The noisy character of the measurement can be well illustrated by considering the marginals of $\tilde{p} (x,y)$  
\begin{equation}
\label{mtpX}
    \tilde{p}_X(x) = \int dy \tilde{p} (x,y) = \int dx^\prime p_X(x^\prime )W_{0X} (x^\prime - x ) 
\end{equation}
where $p_X(x^\prime )$ is the exact probability distribution of quadrature $X$ for the system state, that can be readily obtained from its Wigner function
\begin{equation}
    p_X(x ) = \int dy W_S (x,y) , 
\end{equation}
and 
\begin{equation}
\label{Kx}
W_{0X} (x ) = \frac{1}{\sqrt{2\pi\Sigma_{xx}}} \exp \left[
-\frac{x^2}{2 \Sigma_{xx}} \right ] .
\end{equation}
Similarly for the $Y$ quadrature. That is to say that $\tilde{p} (x,y)$ has an additional noise in $X$ and $Y$ increasing its variances in the quantities $\Sigma_{xx}$ and $\Sigma_{yy}$, respectively. We can also notice that in this increase of uncertainty in $X$ and $Y$ there is no direct contribution arising from the measurement-induced correlations $\Sigma_{xy}$, only the indirect effect through the uncertainty relation (\ref{mS}).

Incidentally, Eqs. (\ref{tp}) and (\ref{Kx}) give us a suitable expression for the marginal POVM in the $x$ and $y$ variables  
\begin{equation}
    \Delta_X (x) = \int dy \Delta (x,y) ,
\end{equation}
that is 
\begin{equation}
\label{mXPOVM}
    \Delta_X (x) = \int dx^\prime W_{0X} (x-x^\prime) |x^\prime \rangle \langle x^\prime | ,
\end{equation}
where $|x \rangle $ are the eigenstates of $X$.

\bigskip

Let us consider a definite example in the form of a system state with Gaussian Wigner function 
\begin{equation}
\label{Gc}
W_S (x,y) = \frac{1}{2\pi\sqrt{\det \Gamma}} \exp \left[
-\frac{1}{2} (x,y)\Gamma^{-1} \binom{x}{y} \right] .
\end{equation}
The result for the joint statistics is 
\begin{equation}
\tilde{p}  (x,y) = \frac{1}{2\pi\sqrt{\det \Phi}} \exp \left[
-\frac{1}{2} (x,y)\Phi^{-1} \binom{x}{y} \right] ,
\end{equation}
with $\Phi = \Gamma + \Sigma$, with marginals 
\begin{equation}
\label{prio}
\tilde{p}_Z  (z) = \frac{1}{\sqrt{2\pi \Phi_{zz}}} \exp \left (
- \frac{z^2}{2 \Phi_{zz}} \right ), 
\end{equation}
for $Z=X,Y$, $z=x,y$.

\section{Conditional distribution} 

Let us address the conditional distribution in the Gaussian case (\ref{Gc}) which is defined by the Bayes' rule 
\begin{equation}
    \tilde{p}_{x|y} (x |y) = \frac{\tilde{p}(x,y)}{\tilde{p}_Y(y)} ,
\end{equation}
to get a Gaussian distribution
\begin{equation}
\label{post}
\tilde{p}_{x|y} (x|y)= \frac{1}{\sqrt{2\pi\sigma_{x|y}^2}} \exp \left[ -\frac{(x-\mu_{x|y})^2} {2\sigma_{x|y}^2}
\right],
\end{equation}
with 
\begin{equation}
\label{ms}
\mu_{x|y}=\frac{\Phi_{xy}}{\Phi_{yy}}y, \qquad   \sigma_{x|y}^2 =\frac{\Phi_{xx} \Phi_{yy}-\Phi^2_{xy}}{\Phi_{yy}}  ,
\end{equation}
where we can see that $\sigma_{x|y}$ does not depend on $y$. To some extent $\tilde{p}_{x|y} (x |y)$ expresses the remaining information on $X$ once we know the value that takes $Y$. In quantum mechanics the only way to say that we know the $Y$ value is to measure $Y$, otherwise we may even say that it does not exist.  So $\tilde{p}_{x|y} (x|y)$ points to a kind of reduced state after a $Y$ measurement, which is nevertheless different from the usual idea of quantum collapse \cite{EW63,MO97,GA00,MO03,MS05,FG14,BLSSU13}. Let us elaborate on this.

Conditioning implies a change on the statistics of $X$ from the prior $\tilde{p}_X (x)$ in Eq. (\ref{prio}) to the posterior $\tilde{p}_{x|y} (x|y)$ in Eq. (\ref{post}). However, typical rules of state reduction do no apply to account for this change, as clearly revealed by the fact that the posterior $\tilde{p}_{x|y} (x|y)$ does depend on $\Sigma_{xy}$ while $\Delta_Y (y)$ does not. In any case, following the Gleason-Born formulation of quantum statistics, we may ask whether there is a reduced state $\rho_y$ such that it reproduces the $x$-statistics represented by $\tilde{p}(x | y)$ for fixed $y$, that is
\begin{equation}
\label{red}
    \tilde{p}_{x|y} (x|y) = \mathrm{tr} \left [\rho_y \Delta_X (x) \right ] ,
\end{equation}
where $\Delta_X (x)$ is the marginal POVM for $X$ in Eq. (\ref{mXPOVM}). The answer in general is negative. In particular we can easily find examples within the Gaussian domain. To this end we note that relations (\ref{mtpX}) and (\ref{Kx}) imply that the $x$ variance of $\tilde{p}_{x|y} (x|y)$ must be larger than the variance of the Kernel (\ref{Kx}), that is 
\begin{equation}
\label{cond}
    \sigma_{x|y}^2 \geq \Sigma_{xx} .
\end{equation}
The possible violation of this condition can well illustrated with the case of the system state being in vacuum where 
\begin{equation}
    \Gamma = \frac{1}{4} \begin{pmatrix} 1&0\\ 0 & 1 \end{pmatrix} 
\end{equation}
in such a way that 
\begin{equation}
\Phi_{xx} = \frac{1}{4} + \Sigma_{xx}, \quad \Phi_{yy} = \frac{1}{4} + \Sigma_{yy}, \quad \Phi_{xy} = \Sigma_{xy} ,
\end{equation}
so that condition (\ref{cond}) becomes
\begin{equation}
    \Sigma_{xy}^2 \leq \frac{1}{4} \left ( \frac{1}{4} + \Sigma_{yy} \right ) .
\end{equation}
Looking at the expressions (\ref{S}), it is easy to see that whenever $r \neq 0$ and $\theta \neq 0,\pi/2$ it is possible to find values of $r$ and $\theta$ so that this condition is not satisfied. For example if $\theta=\pi/4$ we have that the condition (\ref{cond}) is violated for all squeezing parameters with $|r| \geq 0.6585$. Therefore there cannot be a physical $\rho_y$ satisfying (\ref{red}).

\section{Minimizing conditioned variance} 

Given the above result it seems interesting to look for the conditions under which the conditional distribution $\tilde{p} (x|y)$ has minimum uncertainty. For definiteness we focus on Gaussian system states with arbitrary covariance matrix $\Gamma$. Moreover, let us consider pure states so that $\Gamma_{xx} \Gamma_{yy} - \Gamma_{xy}^2=1/16$. The goal is to minimize $\sigma_{x|y}$ by varying $\Gamma$. 

To begin with we can express $\Gamma_{xx}$ in terms of the other $\Gamma$ variables in Eq. (\ref{ms}), and then take variations on $\Gamma_{xy}$, for example, to find a minimum when 
\begin{equation}
    \Gamma_{xy} = \Gamma_{yy}\frac{\Sigma_{xy}}{\Sigma_{yy}} ,
\end{equation}
that leads after a long but straightforward algebra to
\begin{equation}
    \sigma_{x|y}^2 = \frac{1}{16} \left ( \frac{1}{\Sigma_{yy}} + \frac{1}{\Gamma_{yy}} \right ) ,
\end{equation}
where we have used Eq. (\ref{mS}). Then, we get that there is an infimum for $\sigma_{x|y}$ that is approached in the limit $\Gamma_{yy} \rightarrow \infty$ 
\begin{equation}
\label{sS}
    \sigma_{x|y \, {\rm inf}}^2 = \frac{1}{16 \Sigma_{yy}} = \Sigma_{xx} - \frac{\Sigma_{xy}^2}{\Sigma_{yy}} .
\end{equation}
In this limit the input state tends to be arbitrarily compressed in the following quadrature 
\begin{equation}
    X - \frac{\Sigma_{xy}}{\Sigma_{yy}} Y .
\end{equation}
This can be easily seen by noting that 
\begin{equation}
\Delta^2 \left (  X - \frac{\Sigma_{xy}}{\Sigma_{yy}} Y \right ) = \frac{1}{16 \Gamma_{yy}} \rightarrow 0. 
\end{equation}

\bigskip

It is worth noting that the minimum  $\sigma_{x|y}$ is obtained for an input state with $\Delta X \rightarrow \infty$. To some extent this defies common quantum intuition. This is because one would expect that $Y$ conditioning would lead to an increase of $X$ fluctuations, if we rely on the parallels between $Y$ conditioning and $Y$ measurement. However we have shown that there are situations where the opposite holds. We can also notice that the intrinsic noise introduced by the measurement process in quadrature $X$ is $\Sigma_{xx}$ while the infimum  $\sigma_{x|y}^2$ is clearly below, as shown in Eq. (\ref{sS}), pointing to the nonexistence of the reduced $\rho_y$ as commented above. 

We stress again the importance for this result of a nonvanishing correlation $\Sigma_{xy} \neq 0$ introduced by the detection process. Within the quantum regime these measurement induced correlation may be ascribed to the nonclassical nature of the reference state, the squeezed vacuum, which implies the nonclassical nature of the measurement \cite{LA20}. Among other consequences we have that this will produce entanglement in the mode fields leaving the beam splitter introducing nonclassical correlations between the measured quadratures $X_1$, $Y_2$.  

\section{Practical realization}

The POVM (\ref{POVM}) in the form of squeezed Q function we are dealing with can be easily implemented experimentally via double homodyne detection \cite{AK65,CLG87,LP93a,LP93b,KL08}. A possible realization deals with a lossless beam splitter mixing the signal beam with a reference beam in an squeezed vacuum state. A slightly different alternative was considered in Ref. \cite{LM17}. At one of the output beams a quadrature is measured, say $X_1$, via homodyne detection with a sufficiently strong local oscillator. At the other output beam another quadrature is measured, say $Y_2$ by the same procedure. Assuming real transmission and reflection coefficients and a 50 \% beam splitting the relation between input and measured quadrature operators is 
\begin{equation}
    X_1 = \frac{1}{\sqrt{2}} \left ( X +  X_0 \right ) , \quad Y_2 = \frac{1}{\sqrt{2}} \left ( Y - Y_0 \right )
\end{equation}
where $X, Y$ refer to the corresponding quadrature operators of the input signal beam, and $X_0,Y_0$ are the quadrature operators for the squeezed vacuum field. Correcting for the signal-beam attenuation at the beam splitter, we may consider the scaled quadratures $\sqrt{2} X_1$, $\sqrt{2}Y_2$ as the relevant variables 
\begin{equation}
    x= \sqrt{2} X_1 = X +  X_0, \quad  y = \sqrt{2} Y_2 = Y -  Y_0 .
    \end{equation}
The corresponding probability distribution can be computed as the mean value of deltas in the following form
\begin{equation}
\label{mv}
   \tilde{p} (x,y) = \langle \delta (x-X-X_0 ) \delta (y-Y + Y_0 ) \rangle ,
\end{equation}
where it should be noted that there is no ordering problem since the measured operators, i. e., $X +  X_0$, $Y -  Y_0$ commute. The mean value (\ref{mv}) may be carried out using Wigner functions, being the Wigner function of the delta the same delta divided by $\pi$ with variables instead of operators. Each delta operator contributes a factor $1/\pi$ to its Wigner function, which is exactly canceled by the factor $\pi^2$ in the two-mode representation of traces by phase space Wigner integrals. That is, 
\begin{eqnarray}
&\tilde{p} (x,y) = \int dx^\prime dy^\prime dx_0 dy_0 W_S (x^\prime,y^\prime ) W^\prime_0 (x_0,y_0) & \nonumber \\ & & \nonumber \\
& \times \delta (x-x^\prime-x_0 ) \delta (y- y^\prime + y_0 ), &
\end{eqnarray}
where we have primed the Wigner function of the reference state $W^\prime_0$ for reasons we will see soon. This leads to 
\begin{equation}
\tilde{p} (x,y) = \int dx^\prime dy^\prime W_S (x^\prime,y^\prime ) W^\prime_0 (x-x^\prime, y^\prime -y) .
\end{equation}
This only differs from Eq. (\ref{tp}) in the sign of the first argument of the Wigner function of the reference beam. But, being this a Gaussian squeezed vacuum, such sign can be easily absorbed by reversing the sign of the angle $\theta$ as it can be easily seen in Eq. (\ref{S}). So $ W^\prime_0 (x-x^\prime, y^\prime -y) =  W_0 (x^\prime - x, y^\prime -y)$ and we get that the probability of this simple measurement is exactly the same one in Eq. (\ref{tp}).

\section{Classical and semiclassical analyses}

It is instructive to consider the classical analysis of this very same arrangement. It is worth noting that it will follow essentially the same steps considered above regarding its practical implementation. This is because the input-output relations between field quadratures through a lossless beam splitter are exactly the same in quantum and classical optics.

We may distinguish two levels of analysis. We refer to a semiclassical case when the squeezed vacuum is replaced by a fluctuating classical field whose quadratures are distributed according to the very same Wigner function $W_0$, suitably understood here as a legitimate probability distribution over the phase space of the system. We refer to a classical analysis when the squeezed vacuum is replaced by nonfluctuating quadratures with null value, that is to say, when understanding vacuum as nothingness.   

In the classical case, after Eq. (\ref{tp}) we have just 
\begin{equation}
    W_{0 \, \mathrm{class}} (x-x^\prime, y -y^\prime) = \delta  (x-x^\prime )\delta ( y -y^\prime) ,
\end{equation}
so that we have naturally that 
\begin{equation}
\tilde{p}_\mathrm{class} (x,y) = W_S (x,y ),
\end{equation}
always in the understanding that in the classical domain the Wigner function represents a legitimate joint probability distribution. This is the proper classical limit in which any two field quadratures can be measured exactly. This reveals also that in the scheme we are considering the apparatus is a quantum device to the letter. This is an idea we have already followed elsewhere \cite{LA20}.

It is also interesting to examine the semiclassical case in which we consider the reference beam to be a fluctuating field whose statistics coincides with the Wigner function of the squeezed vacuum, to be understood here as a legitimate joint probability distribution which is possible whenever it is nonnegative. We may refer to this as semiclassical in the sense of giving physical nature to the vacuum against the classical nothingness. It is known that there is the possibility of explaining some quantum effects within a classical-like scenario by assuming that the vacuum is such physical fluctuating field \cite{ES12,TB19}. This is the case for example of the spontaneous emission. Nevertheless, the squeezed vacuum is a clear example of nonclassical state of light,  and the spontaneous emission a signature of quantum behavior \cite{MW95,SZ97}. 

In this context we can follow exactly the same steps followed above when analyzing the practical implementation to obtain again the same result (\ref{tp}). The only difference is the assumption that the Wigner functions are true probability distribution functions. With this we encounter in this semiclassical domain the same effect of lack of field state able to account for the conditional distribution $\tilde{p}(x|y)$.  

\section{Qubit case}

The question about joint measurements, conditional probabilities and their Born-like formulation were already addressed for the qubit case in Ref. \cite{PL22}. Here we just complete the analysis looking for the conditions leading to the minimum fluctuations for the conditional distribution $\tilde{p}(x|y)$, that is to say looking for the minimum effect of complementarity.  

For the qubit scenario, a suitable POVM representing a joint noisy measurement of the observables represented by operators $\sigma_X$ and $\sigma_Y$ is of the form
\begin{equation}
\Delta(x,y) = \frac14 \left( \sigma_0+ x\gamma_X\sigma_X+ y\gamma_Y\sigma_Y + xy\gamma_{XY}\sigma_Z \right),
\end{equation}
where $x,y$ are dichotomic variables $x,y=\pm 1$, $\sigma_{X,Y,Z}$ are the Pauli matrices, $\sigma_0$ is the $2\times2$ identity matrix, and the $\gamma$ factors express the noisy character of the measurement with 
\begin{equation}
    \gamma_X^2 + \gamma_Y^2 + \gamma_{XY}^2 \leq 1, 
\end{equation}
so that $\Delta(x,y) \geq 0$. The system state can be as well expressed as
\begin{equation}
\rho= \frac12 (\sigma_0+\bm{s}\cdot\boldsymbol{\sigma}),
\end{equation}
where $\bm{s}$ is a three-dimensional real vector with modulus bounded by unity,
\begin{equation}
\bm{s}=(s_X,s_Y,s_Z), \qquad |\bm{s}|\le 1.
\end{equation}

With all this the joint distribution is 
\begin{equation}
\tilde{p}(x,y) = \frac14 \left( 1+x\gamma_Xs_X+y\gamma_Ys_Y+xy\gamma_{XY}s_Z \right),
\end{equation}
and the $Y$ marginal is 
\begin{equation}
\tilde{p}_Y(y) = \frac12(1+y\gamma_Ys_Y),
\end{equation}
so that the conditional probability is 
\begin{equation}
\label{cpqc}
\tilde{p}(x|y) = \frac12 \left( 1+ \frac{x\gamma_Xs_X+xy\gamma_{XY}s_Z} {1+y\gamma_Ys_Y} \right).
\end{equation}

For the case of dichotomic variables $x,y=\pm 1$ minimum variance is equivalent to larger mean value. In turn, this is equivalent to larger probability. Then in the spirit of the previous analyses let us look for the maximum of $\tilde{p}(x|y)$. In order to maximize $\tilde{p}(x|y)$ we focus first on the numerator of Eq. (\ref{cpqc}) noting that it is the scalar product of two two-dimensional vectors $(x\gamma_X,xy\gamma_{XY})$ and $(s_X,s_Z)$. This is bounded from above for the optimum case which holds when both vectors are proportional 
\begin{equation}
    s_X = \frac{x \gamma_X}{\sqrt{\gamma_X^2 + \gamma_{XY}^2}} \mu , \quad s_Z = \frac{xy\gamma_{XY}}{\sqrt{\gamma_X^2 + \gamma_{XY}^2}} \mu
\end{equation}
where $\mu \leq \sqrt{1-s_Y^2}$ 
so that
\begin{equation}
x\gamma_Xs_X+xy\gamma_{XY}s_Z \le \sqrt{\gamma_X^2+\gamma_{XY}^2}\sqrt{1-s_Y^2}.
\end{equation}
Then, taking variations with respect $s_Y$ we finally get
\begin{equation}
p_{\max}(x|y) = \frac12 \left( 1+ \frac{\sqrt{\gamma_X^2+\gamma_{XY}^2}} {\sqrt{1-\gamma_Y^2}} \right).
\end{equation}
The state reaching the maximum is pure with 
\begin{equation}
s_X = x \gamma_X \, \sqrt{\frac{1-\gamma_Y^2}{\gamma_X^2+\gamma_{XY}^2}}, \quad 
s_Y=-y\gamma_Y,
\end{equation}
and 
\begin{equation}
s_Z  = xy\gamma_{XY} \sqrt{\frac{1-\gamma_Y^2}{\gamma_X^2+\gamma_{XY}^2}} .
\end{equation}
Finally, if the POVM is optimal in the sense that 
\begin{equation}
\label{opt}
    \gamma_X^2 + \gamma_Y^2 + \gamma_{XY}^2 = 1 ,
\end{equation}
we get, maybe surprisingly, that 
\begin{equation}
\tilde{p}_{\max}(x|y)=1 ,
\end{equation}
which holds for the system state 
\begin{equation}
\label{max}
    s_X = x \gamma_X, \quad s_Y = - y \gamma_Y, \quad s_Z = x y \gamma_{XY} .
\end{equation}
We can stress again the paradoxical appearance of the result. This is that $Y$ conditioning can lead to an exact determination of the conjugate variable $X$.

Here again it can be easily seen that whenever $\gamma_X \neq 1$ there is no physical state $\rho_y$ that can reproduce this conditional probability through the marginal POVM, 
\begin{equation}
\tilde{p} (x|y)=\mathrm{tr}[\rho_y\Delta_X(x)],
\end{equation}
where 
\begin{equation}
    \Delta_X (x) = \frac{1}{2} \left ( \sigma_0 + x \gamma_X \sigma_X \right ) .
\end{equation}
This has been further examined in Ref. \cite{PL22}.

\section{Conclusions}

We have examined conditional probabilities derived from the joint noisy measurement of two complementary observables in two interesting quantum-optical scenarios. In both cases we have shown that conditioning can lead to reduction of uncertainty, or even a complete removal. This is accompanied by the lack of a Born-like form for the conditional statistics. This is quite interesting if we consider that conditioning might be regarded as the effect of measuring the complementary variable. But such induced transformation of the statistics does not follow standard rules of state reduction associated to measurement, that otherwise would imply an increase of uncertainty for the posterior distribution.

\end{document}